\documentclass[preprint,11pt]{JHEP3}

\usepackage{epsfig,multicol,amsmath}

\usepackage{bm}  % for the command \bm (boldface greek letters)
\usepackage{amsmath}     % defines the command \dfrac (fractions in displaystyle)

\usepackage{graphicx}
\usepackage{rotating}
\usepackage{pifont}
\usepackage{relsize}
\usepackage{mathtools}
\usepackage{cite}

\def\beq{\begin{equation}}
\def\eeq{\end{equation}}
\def\bea{\begin{eqnarray}}
\def\eea{\end{eqnarray}}

\def\eq#1{{Eq.~(\ref{#1})}}
\def\fig#1{{Fig.~\ref{#1}}}
\newcommand{\bas}{\bar{\alpha}_S}

\newcommand{\Lb}{\left(}
\newcommand{\Rb}{\right)}
\newcommand{\h}{\frac{1}{2}}

\parskip=\itemsep               %?

\newcommand{\nn}{\nonumber}

\newcommand{\ga}{\gamma}

\title{ Non linear evolution: revisiting the solution in the saturation region}
\author{{\Large Carlos Contreras${}^{a}$,  Eugene Levin${}^{a,b}$ and Rodrigo Meneses${}^{c}$}
\\
${}^a$\,\, Departamento de F\'\i sica, Universidad T\'ecnica
Federico Santa Mar\'\i a \, and  \,Centro Cient\'ifico-Tecnol$\acute{o}$gico de Valpara\'\i so,  Avda. Espa\~na 1680, Casilla 110-V,  Valpara\'\i so, Chile\\
${}^b$ \, Department of Particle Physics, School of Physics and Astronomy,
Tel Aviv University, Tel Aviv, 69978, Israel\\
${}^c$\,\, Escuela de Ingenier\'\i a Civil, Facultad de Ingenier\'\i a, Universidad de Valpara\'\i so, Avda Errazuriz 1834, Valpara\'\i so, Chile\\}

\abstract{ In this paper we revisit the problem of the solution to Balitsky-Kovchegov equation deeply in the saturation domain. We find  that solution  has the form  given in Ref.\cite{LT} but it depends on variable $\bar{z} = \ln(r^2 Q^2_s) + \mbox{Const}$ and the value of $\mbox{Const}$  is calculated in this paper.  We propose the solution for full BFKL kernel  at large $z$ in  the entire kinematic region that satisfies the McLerram-Venugopalan \cite{MV} initial condition.}

\keywords{ BFKL equation, BK  non-linear evolution equation,  dipole approach}
\dedicated{PACS: 13.85.-t, 13.85.Hd, 11.55.-m, 11.55.Bq}
\preprint{TAUP  2990/14\,\, \\
USM-TH-325\\
\today}
\begin{document}
\maketitle
%%%%%%%%%%%%%%%%%%%%%%%%%%%%%%%%%%%%%%%%%%%%%%%%%%%%%%%%%%
\section{Introduction}

%%%%%%%%%%%%%%%%%%%%%%%%%%%%%%%%%%%%%%%%%%%%%%%%%%%%%%%%%%
High energy QCD has reached a mature stage\cite{GLR,MUQI,MUCD,MV,KLBOOK}  and has become the common language to discuss high energy scattering where the dense system of partons (quarks and gluons) is produced.
 The most theoretical progress has been reached in the description of  dilute-dense scattering.  The deep inelastic scattering of electron is well known example of such process.
  For these processes   the non-linear equations that govern such processes,  have been derived and discussed in details \cite{BK,JIMWLK}. The extended phenomenology has been developed based on these equations\footnote{We refer the recent review (see Ref. \cite{REV}) which, in our opinion, gives both: the up-to-date status report on the theoretical development and the discussion of the phenomenological description of the experimental data in CGC/saturation approach.} which describes the main features of the high energy scattering.
For phenomenology the numerical solution to the non-linear equations have been used but  it is important to mention that in two limited cases: deeply in the saturation region \cite{LT}  and in the vicinity of the saturation scale\cite{MUT,IIM}; the analytical solutions have been suggested (see Ref.\cite{IIMU} where the procedure to incorporate these analytical solutions are suggested that leads to successful description of HERA data).

In this short paper we re-visit the solution  deeply in the saturation region\cite{LT}. We have two motivations for this. First,  in the semi-classical approach\cite{BKL} we obtain a different solution with the geometric scaling behaviour\cite{GS}  than in Ref.\cite{LT}. Second, the solution for heavy  ions has not been found for the general BFKL kernel\cite{BFKL} in spite of several attempts to find it (see Refs.\cite{LTHI,KLT}).

 We start by recalling the derivation of Ref. \cite{LT}. The non-linear Baslitsky-Kovchegov equation \cite{BK}  takes the form
 \beq \label{BK}
\frac{\partial N_{01}}{\partial Y}\,=\,\bas\int \frac{d^2x_{02}}{2 \pi} \frac{ x^2_{01}}{x^2_{02}\,x^2_{12}}\Big\{ N_{02} + N_{12} - N_{02}N_{12} - N_{01}\Big\}
\eeq
where $N_{ik}=N\Lb Y, x_{ik},b\Rb$. In \eq{BK} we assume that $b \,\gg\,x_{ik}$. Introducing $N_{ik} \,\,=\,\,1\,\,-\,\,\Delta_{ik}$ we obtain the following equation for $\Delta_{ik}$:
\beq \label{BK1}
\frac{\partial \Delta_{01}}{\partial Y}\,=\,\bas \int \frac{d^2x_{02} }{2 \pi}\, \frac{x^2_{01}}{x^2_{02}\,x^2_{12}}
\Big\{  \Delta_{02}\Delta_{12} \,-\,\Delta_{01} \Big\}
\eeq

Deeply in  the saturation region $ x^2_{01}\,Q_s^2\Lb Y, b\Rb\,\,\gg\,\,1$ where $Q^2_s\Lb Y, b\Rb$ is the new scale:  saturation momentum. It is equal to (see Refs.\cite{GLR,MUT,MUPE})

\beq \label{QS}
Q^2_s\Lb Y, b\Rb\,\,=\,\,Q^2_s\Lb Y=0, b\Rb \,e^{\bas\,\kappa Y}\,\,~~\mbox{with}~~~\,\,\,\kappa\,=\,\,\,\frac{\chi\Lb \gamma_{cr}\Rb}{ 1 - \gamma_{cr}}
\eeq
In \eq{QS}
\beq \label{GACR}
\frac{\chi\Lb \gamma_{cr}\Rb}{1 - \gamma_{cr}}\,\,=\,\, - \frac{d \chi\Lb \gamma_{cr}\Rb}{d \gamma_{cr}}~~~\,\,\,\mbox{and}\,\,\,~~~\chi\Lb \gamma\Rb\,=\,\,2\,\psi\Lb 1 \Rb\,-\,\psi\Lb \gamma\Rb\,-\,\psi\Lb 1 - \gamma\Rb
\eeq
$\chi\Lb \gamma\Rb$ is the kernel of the BFKL linear equation \cite{BFKL} where $\psi(\gamma)\, =\, d \ln \Gamma\Lb \gamma\Rb/d \gamma$  is the Euler psi-function (see  formula {\bf 8.36} of Ref.\cite{RY}).

Assuming that both $x_{12}$ and $x_{02}$ are in the saturation region, i.e. $x^2_{12}\,Q^2_s\Lb Y, b\Rb \,>\,1$ and $
x^2_{02}\,Q^2_s\Lb Y, b\Rb \,>\,1$  we can consider that $\Delta_{ik} \,\ll\,1$ and neglect the term proportional to $\Delta_{02}\,\Delta_{12}$ in comparison with $\Delta_{01}$. Resulting equation takes the form
\bea \label{BKLT}
&&\frac{\partial \Delta_{01}}{\partial Y}\,=\nn \\
&&\,- \bas \Delta_{01} \int \frac{d^2\,x_{02} }{2 \pi}\, \frac{ x^2_{01}}{x^2_{02}\,x^2_{12}}\,\,=\,\,- \frac{\bas}{2 \pi} \Delta_{01}\Bigg\{\underbrace{ \pi \int^{x_{01}^2}_{1/Q^2_s} \frac{d x^2_{02}}{x^2_{02}}}_{ \mbox{ $x_{02} \ll  x_{01}$}} \,\,\,+\,\,\,\underbrace{\,\pi  \int\int^{x_{01}^2}_{1/Q^2_s}  \frac{d x^2_{12}}{x^2_{12}} }_{\mbox{ $x_{12} \ll  x_{01}$}} \Bigg\}\ = - \bas  z \Delta_{01}
\eea
where we introduce a new variable
\beq \label{z}
z\,\,=\,\,\ln\Lb x^2_{01}\,Q^2_s\Lb Y, b\Rb\Rb\,\,\, =\,\,\,\,\bas\,\kappa \,Y\,\,+\,\,\xi\,
\eeq
with  $ \xi\,\,=\,\,\ln\Lb x^2_{01} Q^2_s\Lb Y=0,b\Rb\Rb$.

One can see that the solution to \eq{BKLT}  is
\beq \label{BKLT1}
\Delta_{01}\,\,\,=\,\,\mbox{Const}\,\exp\Big(  - \frac{z^2}{2\,\kappa}\Big)
\eeq
It should be stressed that this solution shows the geometric scaling behaviour\cite{GS}  being function of only one variable: $z$.

This derivation shows two problems that have been mentioned above: we need to assume that the main contribution in \eq{BKLT} stems from the saturation region; and the answer has a geometric scaling behaviour that contradicts the initial condition for the DIS with nuclei.

%%%%%%%%%%%%%%%%%%%%%%%%%%%%%%%%%%%%%%%%%%%%%%%%%%%%%%%%
 \begin{figure}
 \begin{center}
  \leavevmode
      \includegraphics[width=8cm]{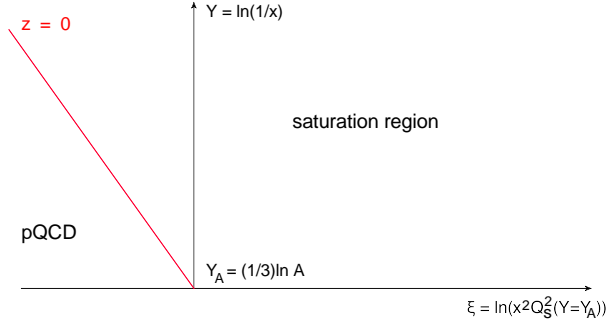}
      \end{center}
      \caption{  Saturation region of QCD. Red line shows the saturation boundary (z=0).}
      \label{sat}
\end{figure}
 %%%%%%%%%%%%%%%%%%%%%%%%%%%%%%%%%%%%%%%%%%%%%%%%%%%%%%%%%

Indeed, at $Y=Y_A$  for DIS with nuclei we have McLerran-Venugopalan formula for the imaginary part of the dipole-nucleus amplitude, which takes the following form (see \fig{sat})
\beq \label{MVF}
N\Lb x^2, Y = Y_A\Rb\,\,\,=\,\,1\,\,\,-\,\,\,\exp\Big( - x^2\,Q^2_s\Lb Y=Y_A,b\Rb/4\Big)\,\,=\,\,1\,\,\,-\,\,\exp\Lb - e^\xi\Rb
\eeq
One can see that \eq{MVF} does not reproduce the solution of \eq{BKLT1} at $Y = Y_A$.  Comparing \eq{MVF} and \eq{BKLT1} we see that the geometric scaling behaviour  cannot be correct in the entire saturation region.

%%%%%%%%%%%%%%%%%%%%%%%%%%%%%%%%%%%%%%%%%%%%%%%%%%%%%%%%
\section{Equation and solution in the momentum space}
%%%%%%%%%%%%%%%%%%%%%%%%%%%%%%%%%%%%%%%%%%%%%%%%%%%%%%%%%
\subsection{Equation and geometric scaling solution}
%%%%%%%%%%%%%%%%%%%%%%%%%%%%%%%%%%%%%%%%%%%%%%%%%%%%%%%%%
We re-write the Balitsky-Kovchegov equation of \eq{BK}  in the momentum space introducing
\beq \label{MR}
N\Lb x^2, b; Y\Rb\,\,=\,\,x^2 \,\int \frac{d^2 k_\perp}{ 2 \pi}\,e^{ i \vec{k}_\perp \cdot \vec{x}}\,\widetilde{N}\Lb k_\perp, b; Y\Rb
\eeq
It takes the form\cite{GLR,KOV}
\beq \label{BKMR}
\frac{\partial \widetilde{N}\Lb k_\perp, b; Y\Rb}{\partial Y}\,\,=\,\,\bas \Bigg\{ \chi\Lb -\,\frac{\partial}{\partial \tilde{\xi}}\Rb
 \widetilde{N}\Lb k_\perp, b; Y\Rb\,\,\,-\,\, \widetilde{N}^2\Lb k_\perp, b; Y\Rb\Bigg\}
 \eeq
 where
 \beq \label{XIT}
 \tilde{\xi}\,\,=\,\,\,\,\ln\Lb k^2_\perp/Q^2_s\Lb Y = Y_A, b\Rb\Rb~~~\mbox{and}~~~\tilde{z}\,\,=\,\,\bas \kappa \Lb Y - Y_A\Rb \,\,-\,\,\tilde{\xi}\,\,=\,\,\ln \Lb Q^2_s\Lb Y, b\Rb/k^2_\perp\Rb
 \eeq

 The advantage of the non-linear equation in \eq{BKMR} that the non-linear term depends only on external variable  and does not contain the integration over momenta.  The  BFKL kernel: $  \chi\Lb -\,\frac{\partial}{\partial \tilde{\xi}}\Rb$, 
 can be written as the series over positive powers of $\partial/\partial \tilde{\xi}$ except of the first term

 \beq \label{FIT}
 \frac{1}{\gamma}\, \widetilde{N}\Lb k_\perp, b; Y\Rb\,\,\, \to\,\,\, \int^{k^2_\perp}\frac{ d k'^2_\perp}{k'^2_\perp}  \widetilde{N}\Lb k'_\perp, b; Y\Rb
 \eeq

 Differentiating \eq{BKMR} over $\tilde{\xi}$ one can see that  it can be re-written as
 \bea \label{BKMR1}
 &&\frac{\partial \widetilde{N}'_{\tilde{\xi}}\Lb k_\perp, b; Y\Rb}{\partial Y} \,\,\,=\\
&& \,\,\,\bas\Bigg\{ \Big(\chi\Lb\gamma\Rb\,\,-\frac{1}{\gamma}\Big)  \widetilde{N}'_{\tilde{\xi}}\Lb k_\perp, b; Y\Rb\,\,+\,\,  \widetilde{N}\Lb k_\perp, b; Y\Rb\,\,-\,\,2  \widetilde{N}'_{\tilde{\xi}}\Lb k_\perp, b; Y\Rb  \widetilde{N}\Lb k_\perp, b; Y\Rb \Bigg\}\nn
 \eea
 where $\gamma\,=\, -\,\frac{\partial}{\partial \tilde{\xi}}$.

  Introducing the variable $\tilde{z}$ instead of $\tilde{\xi}$ and the new function $M$ as
 \beq \label{M}
\widetilde{ N}'_{\tilde{z}}\Lb \tilde{z},b; Y\Rb\,\,=\,\,\h +ÊM\Lb\tilde{z}, b; Y\Rb ~~~~\mbox{or}~~~~\widetilde{N}\Lb  \tilde{z},b; Y\Rb\,\,=\,\,\h \tilde{z}\,\,+\,\,\int^{\tilde{z}}_0 d \tilde{z}'  M\Lb\tilde{z}',b; Y\Rb
 \eeq
 we can re-write \eq{BKMR1} in the form
 \bea \label{BKMR2}
&&\kappa\,\frac{\partial M\Lb\tilde{z}, b; Y\Rb}{\partial \tilde{z}}\,\,+\,\,\frac{\partial M\Lb\tilde{z}, b; Y\Rb}{\partial Y} \,\,\,=\\
&&\bas\Bigg\{ \Big(\chi\Lb \gamma\Rb\,\,-\,\,\frac{1}{\gamma}\Big) M\Lb \tilde{z},  b; Y\Rb \,\,-\,\,\,\tilde{z}\,M\Lb \tilde{z}, b; Y\Rb\,\,-\,\,M\Lb\tilde{z}, b; Y\Rb \int^{\tilde{z}}_0 d \tilde{z}'  M\Lb\tilde{z}',b; Y\Rb \Bigg\}\nn
\eea
 with $\gamma = \frac{\partial}{\partial \tilde{z}}$.

 We  are going to find solution inside the saturation region where function $M$ is small at large $\tilde{z}$.
 However, we need to re-write \eq{BKMR2} replacing it by
  \bea \label{BKMR21}
&&\kappa\,\frac{\partial M\Lb\tilde{z}, b; Y\Rb}{\partial \tilde{z}}\,\,+\,\,\frac{\partial M\Lb\tilde{z}, b; Y\Rb}{\partial Y} \,\,\,=\\
&&\bas\Bigg\{ \Big(\chi\Lb \gamma\Rb\,\,-\,\,\frac{1}{\gamma}\Big) M\Lb \tilde{z},  b; Y\Rb \,\,-\,\,\,\Lb \tilde{z}\,+\,\lambda\Rb\,M\Lb \tilde{z}, b; Y\Rb\,\,+\,\,M\Lb\tilde{z}, b; Y\Rb \int^\infty_{\tilde{z}}d \tilde{z}'  M\Lb\tilde{z}',b; Y\Rb \Bigg\}\nn
\eea
where
\beq \label{LAMBDA}
\lambda\,\,=\,\,\int^\infty_0 \,d \tilde{z}'  M\Lb\tilde{z}',b; Y\Rb
\eeq
and neglecting  the last term in this equation one can  see that we need to solve the following  linear equation
 \beq \label{BKMR3}
\kappa\,\frac{\partial M\Lb\tilde{z}, b; Y\Rb}{\partial \tilde{z}}\,\,+\,\,\frac{\partial M\Lb\bar{z}, b; Y\Rb}{\partial Y} \,\,\,=\
\bas\Bigg\{ \Big(\chi\Lb \gamma\Rb\,\,-\,\,\frac{1}{\gamma}\Big) M\Lb \bar{z},  b; Y\Rb \,\,-\,\,\,\bar{z}\,M\Lb \tilde{z}, b; Y\Rb\,\, \Bigg\}
\eeq
with  $\gamma\,\,=\,\,\frac{\partial}{\partial \bar{z}} \,\,\mbox{and}\,\,\,\bar{z}\,\,=\,\,\tilde{z}\,+\,\lambda$

First we find the geometrical scaling solution which depends only on $\tilde{z}$. In this case \eq{BKMR3} takes the form
\beq \label{BKMR4}
\kappa \frac{ d M\Lb\bar{z}, b\Rb}{d \bar{z}}\,\,\,=\,\,\,\Big(\chi\Lb \,\gamma\Rb\,\,-\,\,\frac{1}{\gamma}\Big) M\Lb \bar{z},  b\Rb \,\,-\,\,\,\bar{z}\,M\Lb \bar{z}, b\Rb
\eeq

The boundary  condition for this equation we take
\beq \label{ICGS}
 N'_{\tilde{z}}\Lb \tilde{z}\,=\,0,b\Rb\,\,=\,\,\h +ÊM\Lb\tilde{z}\,=\,0, b\Rb\,\,=\,\,\phi_0(b)\,\, \ll\,\,1
\eeq
where $\phi_0(b)$ is the solution to the linear BFKL equation at $\tilde{z}=0$. $\phi_0(b)\,\leq\,1$  
due to unitarity constraint and should be small to neglect that non-liner term at $\tilde{z} = 0$.

\eq{BKMR4} can be solved using the Mellin transform
\beq \label{MT}
M \Lb\tilde{z}, b\Rb= \int^{\epsilon + i \infty}_{\epsilon - i \infty} \frac{d \gamma}{2 \pi i} e^{\gamma \,\bar{z}} m\Lb \gamma, b\Rb
\eeq
where  $m\Lb \gamma, b\Rb$ satisfies the equation:
  \beq \label{BKMT1}
 \Lb \kappa \gamma - \chi\Lb \gamma\Rb + 1/\gamma\Rb\,m\Lb \gamma,b\Rb = \frac{d m\Lb \gamma, b\Rb}{d \gamma}.
  \eeq

The solution for $m\Lb \gamma, b\Rb$ takes the following form
   \beq \label{BKMT2}
   m\Lb \gamma\Rb\,\,=\,\,\exp\Bigg(\int^\gamma_0  d \gamma' \Big(\kappa \gamma' -  \chi\Lb \gamma'\Rb +1/\gamma' \Big) \Bigg)
   \eeq
and taking into account the explicit form of the BFKL kernel given by \eq{GACR} one can re-write \eq{BKMT2} in the form
      \beq \label{BKMT3}
    m\Lb \gamma\Rb\,\,=\,\,\exp\Bigg(\kappa \gamma^2/2 -  2 \psi(1)\gamma  \Bigg)\,\Bigg(\frac{\gamma\Gamma( \gamma)  }{ \Gamma(1 - \gamma)}\Bigg) \,\,=\,\,\exp\Bigg(\kappa \gamma^2/2 -  2 \psi(1)\gamma  \Bigg)\,\frac{\Gamma(1 +  \gamma)  }{ \Gamma(1 - \gamma)}
        \eeq

 Substituting \eq{BKMT3} into \eq{MT} we obtain
 \beq \label{BKMT4}
 M \Lb\bar{z}, b\Rb= \int^{\epsilon + i \infty}_{\epsilon - i \infty} \frac{d \gamma}{2 \pi i} e^{\gamma \bar{z}\,+\,\kappa \gamma^2/2}\,\frac{\Gamma(1 +  \gamma)  }{ \Gamma(1 - \gamma)}
  \eeq
     where  for $\bar{z} $ we use a new definition:  $\bar{z} \,\,=\,\,\tilde{z} \,+\,\lambda \,\,- 2 \,\,\psi\Lb 1 \Rb$.

    One can see that in \eq{BKMT4} we cannot close the contour of integration in $\gamma$ neither on the left semi-plane nor on the right one. Introducing  $\gamma \,=\,i\,\bar{\gamma}$ we reduce \eq{BKMT4} to the form
   \beq \label{BKMT5}
 M \Lb\bar{z}, b\Rb= \int^{+ \infty}_{- \infty} \frac{d \bar{\gamma}}{2 \pi } e^{i\,\bar{\gamma} \bar{z}\,-\,\kappa \bar{\gamma}^2/2}\,\frac{\Gamma(1 + i \bar{\gamma})  }{ \Gamma(1 -  i \bar{\gamma})}
  \eeq

      For large $\tilde{z}$  and  $\bar{\gamma} $ is large about $\tilde{z}$, we can use the approximation
      \beq \label{GG}
     \Gamma\Lb 1 + i \bar{\gamma}\Rb\,\,\xrightarrow{ \bar{\gamma} \,\gg\,1} \,\,\sqrt{2 \pi} | \bar{\gamma}|^{\h}\,e^{ - \h \pi |\bar{\gamma}|}
\eeq
(see formula {\bf 8.328}  of Ref.\cite{RY}).  Using \eq{GG},  \eq{BKMT5}  takes the form
   \beq \label{BKMT51}
 M \Lb\bar{z}, b\Rb\,\,= \,\,\int^{+ \infty}_{- \infty} \frac{d \bar{\gamma}}{2 \pi } e^{i\,\bar{\gamma} \bar{z}\,-\,\kappa \bar{\gamma}^2/2\, }
  \eeq
which is equal (see formulae {\bf 3.462(3), 9.246} of Ref. \cite{RY})
\beq \label{BKMT6}
 M \Lb\bar{z}, b\Rb\,\,=\,\,\frac{1}{\sqrt{2 \pi \kappa}}\, e^{- \frac{\bar{z}^2}{4 \kappa}}\,D_0\Lb -\frac{\bar{z}}{\sqrt{\kappa}}\Rb\,\,\,=\,\,\frac{1}{\sqrt{2 \pi \kappa} }\,e^{ - \frac{\bar{z}^2}{2 \kappa}}
 \eeq
 where   $D_n(z)=  2^{- \h n} e^{- \frac{\bar{z}^2}{4}} H_n(z/\sqrt{2})$ is the parabolic cylinder function (see  formulae {\bf 9.24 - 9.25} of Ref.\cite{RY}).

  Therefore we reproduce the solution of \eq{BKLT1}.   Choosing the coefficient in front of \eq{BKMT6} we can easily satisfy the initial condition of \eq{ICGS} which leads to the solution
  \beq \label{BKMT7}
 M \Lb\bar{z}, b\Rb\,\,=\,\,\Lb \phi_0\Lb b \Rb - \h\Rb\,e^{ - \frac{\bar{z}^2}{2 \kappa}}
 \eeq
 %%%%%%%%%%%%%%%%%%%%%%%%%%%%%%%%%%%%%%%%%%%%%%%%%%%%%%%%%
 \begin{boldmath}
\subsection{General solution and initial condition at $Y = Y_A$}
\end{boldmath}
%%%%%%%%%%%%%%%%%%%%%%%%%%%%%%%%%%%%%%%%%%%%%%%%%%%%%%%%%
 As has been mentioned we are not able to find the geometric scaling solution that satisfy both initial and boundary conditions given by \eq{MVF} and \eq{ICGS}. We need to solve a general \eq{BKMR3} to find such a solution. We start with re-writing boundary condition of \eq{MVF} for function $M \Lb\tilde{z}, b; Y \Rb $ in momentum representation.
 \bea \label{BC}
\h \, \,+\,M\Lb \tilde{ z},b; Y=Y_A\Rb\,\,&=&\,\,\frac{d}{d \tilde{z}}\int \frac{d^2 r}{r^2}\,e^{i \vec{k}\cdot \vec{r}}\,\Big( 1\,-\,\exp\Lb - r^2 \,Q^2_s\Lb b; Y= Y_A\Rb/4\Rb\Big)\,\,\nn\\
&=&\,\,\frac{d}{ d \tilde{z}}\Lb \h\Gamma_0\Lb \frac{k^2_\perp}{Q^2_s\Lb b; Y= Y_A\Rb}\Rb\Rb\,\,=\frac{d}{ d \tilde{z}}\Lb \h\Gamma_0\Lb e^{-\,\tilde{z}}\Rb\Rb\,\,=\,\,\h \exp\Big( -\,e^{-\,\tilde{z}}\Big) \nn\\
 M\Lb  z, b; Y=Y_A\Rb &=&- \h \Lb 1 \,\,-\,\, \exp\Big( -e^{-\,\tilde{z}}\Big)\Rb
  \eea

 We solve \eq{BKMR3} using the double Mellin transform: viz.
 \beq \label{DMT}
    M\Lb \tilde{z}, b; Y\Rb\,\,=\,\,\int^{\epsilon + i \infty}_{\epsilon  - i \infty} \frac{d \omega}{ 2 \pi i} \int^{\epsilon + i \infty}_{\epsilon - i \infty} \frac{d \gamma}{ 2 \pi i}   e^{\omega \Lb Y   -  Y_A\Rb+ \gamma \tilde{z}} \,m\Lb \omega, \gamma; b \Rb
\eeq
For $m\Lb \omega, \gamma; b \Rb$ the equation takes the form\footnote{We omitted argument $b$ in $m\Lb \omega, \gamma; b \Rb$ since our equations do not depend on $b$ and it enters only through the  initial and boundary conditions.}
  \beq \label{BKDM1}
    \Lb \omega \,+ \,\kappa \gamma \,-\, \chi\Lb \gamma\Rb\,+\,\frac{1}{\gamma}\Rb m\Lb  \omega, \gamma\Rb\,=\,\frac{\partial  m \Lb \omega, \gamma\Rb}{\partial \gamma}
\eeq
 Solution to \eq{BKDM1} can be written in the form
\bea \label{BKDM2}
   M\Lb \bar{z},  Y\Rb&=&\int^{\epsilon + i \infty}_{\epsilon - i \infty}\!\! \frac{d \omega}{ 2 \pi i} e^{\omega \Lb Y  -  Y_A\Rb}\int^{\epsilon + i \infty}_{\epsilon -  i \infty}\!\! \frac{d \gamma}{ 2 \pi i} I\Lb \omega\Rb \exp\Bigg( \omega \,\gamma\,\,+\,\,\gamma \bar{z} + \kappa \gamma^2/2 \Bigg)\frac{\Gamma( 1 + \gamma)}{ \Gamma(1 - \gamma)}
       \eea
  where function $I\Lb \omega \Rb$ has to be found from \eq{BC}. At $Y= Y_A$ \eq{BC} can  be written as
   \bea \label{BCFM}
    &&  M\Lb Y = Y_A , \bar{z}\Rb  \,=\nn\\
    &&\,\int^{\epsilon + i \infty}_{\epsilon - i \infty}\!\! \frac{d \omega}{ 2 \pi i}\int^{\epsilon + i \infty}_{\epsilon -  i \infty}\!\! \frac{d \gamma}{ 2 \pi i} I\Lb \omega\Rb \exp\Bigg( \omega \,\gamma\,\,+\,\,\gamma \bar{z} + \kappa \gamma^2/2  \Bigg)\frac{\Gamma( 1 + \gamma)  }{ \Gamma(1 - \gamma)}\nn\\
    &&=\,\int^{\epsilon + i \infty}_{\epsilon -  i \infty}\!\! \frac{d \gamma}{ 2 \pi i} J\Lb \gamma\Rb \exp\Bigg( \gamma z + \kappa \gamma^2/2 \,+\, 2 \psi(1)\gamma \Bigg)\frac{\Gamma( 1 + \gamma)  }{ \Gamma(1 - \gamma)}  \eea
  One can see that
  \beq \label{J}
  J\Lb \gamma \Rb\,\,=\int^{\epsilon + i \infty}_{\epsilon - i \infty}\!\! \frac{d \omega}{ 2 \pi i} \,e^{\omega\,\gamma}\,I\Lb \omega \Rb\,=\,
  \h\,
  \frac{\Gamma\Lb 1 - \gamma\Rb}{\Gamma\Lb 1 + \gamma\Rb } \,\Lb\Gamma\Lb \gamma\Rb\,\, -\,\, \frac{1}{\gamma}\Rb\,\exp\Lb - \frac{\kappa \gamma^2}{2}  \Rb
  \eeq
    leads to \eq{BCFM}. Indeed, substituting \eq{J} into \eq{BCFM} one obtains \eq{BC}  closing contour in $\gamma$ over negative $\gamma$-s

   For $ Y  >  Y_A$ the solution takes the form\footnote{For simplicity we use $Y= Y- Y_A$ to the end of this section. We hope  that using the same letter $Y$ for both variables, will not cause any inconvenience.}
   \bea \label{YSOL}
 &&  M\Lb Y  , \bar{z}\Rb \,\,=\,\,\int^{\epsilon + i \infty}_{\epsilon -  i \infty}\!\! \frac{d \gamma}{ 2 \pi i} J\Lb Y\, \,+\,\gamma\Rb \exp\Bigg( \gamma \bar{z} + \kappa \gamma^2/2 \Bigg)\frac{\Gamma( 1 + \gamma)  }{ \Gamma(1 - \gamma)} \\
   &&=\,\,\,  \h \int^{\epsilon + i \infty}_{\epsilon -  i \infty}\!\! \frac{d \gamma}{ 2 \pi i}\exp\Big(\gamma \bar{z}  - \h \kappa Y^2  - \kappa Y \gamma \Big)\,\frac{\Gamma\Lb 1 + \gamma\Rb}{\Gamma\Lb 1 - \gamma\Rb}\,
    \frac{\Gamma\Lb 1 - \gamma - Y\Rb}{\Gamma\Lb 1 + \gamma + Y\Rb } \,\Lb\Gamma\Lb \gamma + Y\Rb \,\,-\,\, \frac{1}{\gamma + Y}\Rb
   \nn\\
   &&=\,\,\, \h e^{ - \h \kappa Y^2} \int^{\epsilon + i \infty}_{\epsilon -  i \infty}\!\! \frac{d \gamma}{ 2 \pi i}\exp\Big(-\gamma \tilde{\xi}    \,+\,2\,\psi(1)\,Y\Big)\,\frac{\Gamma\Lb 1 + \gamma\Rb}{\Gamma\Lb 1 - \gamma\Rb}
    \frac{\Gamma\Lb 1 - \gamma - Y\Rb}{\Gamma\Lb 1 + \gamma + Y\Rb } \,\Lb\Gamma\Lb \gamma + Y\Rb \,\,-\,\, \frac{1}{\gamma + Y}\Rb \nn\\
    &&=\,\,\,- \h e^{ - \h \kappa Y^2} \int^{\epsilon + i \infty}_{\epsilon -  i \infty}\!\! \frac{d \gamma}{ 2 \pi i}\exp\Big(-\gamma \tilde{\xi}    \,+\,2\,\psi(1)\,Y\Big)\,\frac{\Gamma\Lb 1 + \gamma\Rb}{\Gamma\Lb 1 - \gamma\Rb}
    \frac{\Gamma\Lb  - \gamma - Y\Rb}{\Gamma\Lb  \gamma + Y\Rb } \,\Lb\Gamma\Lb \gamma + Y\Rb \,\,-\,\, \frac{1}{\gamma + Y}\Rb \nn\\
    &&=\,\,\,-\h\,e^{ - \h \kappa Y^2\,\, \,+\,2\,\psi(1)\,Y } \,\Bigg\{\,I_1 \Lb Y, \tilde{\xi}\Rb\,\,-\,\,I_2\Lb Y,  \tilde{\xi}\Rb\,\Bigg\}
    \nn
        \eea
  Since $\tilde{\xi} \,\,<\,\,0$ we can take the integrals over $\gamma$ in $I_1\Lb Y,  \tilde{\xi}\Rb$ and in $I_2\Lb Y, \tilde{\xi}\Rb$ closing contours of integrations on the left semi-plane. In    $I_1\Lb \tilde{\xi}\Rb$ and $I_2\Lb \tilde{\xi}\Rb$ we have two sets of poles: $\gamma \,=\,-n - 1 $ from $\Gamma\Lb 1 + \gamma\Rb$ and $\gamma\,\,=\,\,n - [Y]$  where $[Y]$ is the integer part or floor function of $Y$, from $\Gamma\Lb - \gamma - Y\Rb$. These sets lead to the following contributions to $I_1\Lb Y, \tilde{\xi}\Rb$  and
  to $I_2\Lb Y, \tilde{\xi}\Rb$:

  \bea
  I_1\Lb Y, \tilde{\xi}\Rb\,\,\,&=&\,\,\,    I^1_1\Lb Y, \tilde{\xi}\Rb\,\,+\,\,  I^2_1\Lb Y, \tilde{\xi}\Rb\,; \nn\\
  I^1_1\Lb Y, \tilde{\xi}\Rb\,\,\,&=&\,\,\,\sum^\infty_{n=0}\frac{(-1)^n}{n!}\frac{\Gamma\Lb n + 1 - Y\Rb}{\Gamma\Lb n + 2\Rb}\,e^{(n + 1)\tilde{\xi}}\,\,=\,\,e^{\tilde{\xi}}\,\Gamma\Lb 1 - Y\Rb\,{}_1F_1\Lb 1 - Y, 2, - e^{\tilde{\xi}}\Rb\,; \label{I11}\\
   I^2_1\Lb Y, \tilde{\xi}\Rb\,\,\,&=&\,\,e^{\tilde{\xi}\,Y}\,\sum^{[Y]}_{n=0}\frac{(-1)^n}{n!}\frac{\Gamma\Lb n + 1 - Y\Rb}{\Gamma\Lb 1 - n + Y\Rb}\,e^{-n\,\tilde{\xi}}\,; \label{I12}\\
     I_2\Lb Y, \tilde{\xi}\Rb\,\,\,&=&\,\,\,    I^1_2\Lb Y, \tilde{\xi}\Rb\,\,+\,\,  I^2_2\Lb Y, \tilde{\xi}\Rb\,; \nn\\
         I^1_2\Lb Y, \tilde{\xi}\Rb\,\,\,&=&\,\,\,\sum^\infty_{n=0}\frac{(-1)^n}{n!}\frac{\Gamma\Lb n + 1 - Y\Rb}{\Gamma\Lb n + 2\Rb\,\Gamma\Lb Y - n\Rb}\,e^{(n + 1)\tilde{\xi}}\,\,=\,\,e^{\tilde{\xi}}\,\frac{\Gamma\Lb 1 - Y\Rb}{\Gamma\Lb Y\Rb}\,{}_2F_1\Lb 1 - Y,  -Y, 2, -e^{\tilde{\xi}}\Rb\,; \label{I21}\\  
     I^2_2\Lb Y, \tilde{\xi}\Rb\,\,\,&=&\,\,e^{\tilde{\xi}\,Y}\,\sum^{[Y]}_{n=0}\frac{(-1)^n}{n!}\frac{\Gamma\Lb n + 1 - Y\Rb}{\Gamma\Lb 1 - n + Y\Rb\,\Gamma\Lb n + 1\Rb}\,e^{-\,n\,\tilde{\xi}}\,; \label{I22}
       \eea

  In \eq{I11} - \eq{I22} ${}_1F_1\Lb \alpha, \beta, t\Rb$ is the confluent hypergeometric function (another notation is $\Phi\Lb \alpha, \beta, t\Rb$, see formulae {\bf 9.2} of Ref. \cite{RY}) and  ${}_2F_1\Lb \alpha, \beta, \gamma, t\Rb$  is the hypergeometric function (see formulae {\bf 9.1} of Ref. \cite{RY}).

  For matching of this solution with the solution given by \eq{BKMT7} we need to know the asymptotic behaviour of \eq{I11}-\eq{I22} at large values of $Y$. Using Kummer's transformation: $ {}_1F_1\Lb \alpha, \beta, t\Rb\,\,=\,\,e^t \,{}_1F_1\Lb \beta - \alpha, \beta, - t\Rb$ we can re-write \eq{I11} in the form
  \bea \label{I11LY}
    I^1_1\Lb Y, \tilde{\xi}\Rb\,\,\,&=&\,\,e^{\tilde{\xi}}\,\Gamma\Lb 1 - Y\Rb \exp\Lb - e^{\tilde{\xi}}\Rb\,{}_1F_1\Lb 1 + Y, 2, e^{\tilde{\xi}}\Rb\,\,\\
    &\xrightarrow{Y \,\gg\,1}&  \,\,e^{\tilde{\xi}} \exp\Lb - e^{\tilde{\xi}}\Rb \,\Gamma\Lb 1 - Y\Rb\sum^\infty_{n=0}\frac{Y^n}{n! (n + 1)!} e^{n \tilde{\xi}}\,=\,   e^{\tilde{\xi}} \exp\Lb - e^{\tilde{\xi}}\Rb \frac{\Gamma\Lb 1 - Y\Rb}{2 \sqrt{Y \,e^{\tilde{\xi}}}}
    I_1\Lb  2 \sqrt{Y \,e^{\tilde{\xi}}}\Rb\nn
    \eea
    where $I_1\Lb t \Rb$ is the modified Bessel function of the first kind (see formulae {\bf 8.445 - 8.451} of Ref.\cite{RY}).
    Using their asymptotic behaviour at large values of the argument we obtain
 \beq    \label{I11LY1}
    I^1_1\Lb Y, \tilde{\xi}\Rb\,\,\xrightarrow{Y \gg 1}\,\, e^{\tilde{\xi}} \exp\Lb - e^{\tilde{\xi}}\Rb \,e^{-\, Y \Lb\ln Y - 1\Rb}\frac{1}{\sqrt{2 \pi \Lb 4 Y \exp\Lb\tilde{\xi}\Rb\Rb^3}}\,\exp\Lb 2 \sqrt{Y \exp\Lb\tilde{\xi} \Rb}\Rb
     \eeq

    We replace \eq{I12} by the integral,   i.e.
    \beq \label{I12LY}
       I^2_1\Lb Y, \tilde{\xi}\Rb\,\,\to\,\,e^{\tilde{\xi}\,Y}\int ^Y_0 d t \frac{\Gamma\Lb 1 + t - Y\Rb}{\Gamma\Lb 1 - t +Y\Rb\,\Gamma\Lb 1 + t\Rb}\,\,e^{-\tilde{\xi}\,t \,+ \,i\,\pi\,t}
    \eeq
 The steepest decent method leads to the following contribution
 \beq \label{I12SP}
      I^2_1\Lb Y, \tilde{\xi}\Rb\,\,\to\,\,\,\sqrt{\pi \Lb Y \,e^{\tilde{\xi}}\Rb^{\h}} e^{-Y \ln Y}
      \eeq
  at the saddle point $t_{SP}$ which can be found from the equation
 \beq \label{TSP}
 - \ln t_{SP} + 2 \ln\Lb Y - t_{SP} - 1\Rb\,-\,\tilde{\xi}\,=\,0\,; \,\,\,\,\,\,\,\,\,\,\,\,\,t_{SP} \,=\,Y - \sqrt{Y} e^{\h\,\tilde{\xi}}\,\,\,\,\,\mbox{at}\,\,\,\,Y\,\gg\,1
 \eeq
The large $Y$ behaviour of \eq{I21}  can be obtained using the transformation $_{2}F_{1}(\alpha,\beta,\gamma,t)=(1-t)^{\gamma-\alpha-\beta}{}_{2}F_{1}(\gamma-\alpha,\gamma-\beta,\gamma,t)$ (see formulae {\bf  9.131} of Ref. \cite{RY})
\begin{equation*}
\begin{array}{rcl}
I^{1}_{2}( Y, \tilde{\xi})&=& e^{\tilde{\xi}}\frac{\Gamma(1-Y)}{\Gamma(Y)}{}_{2}F_{1}(1-Y,-Y,2,-e^{\tilde{\xi}})=(1+e^{\tilde{\xi}})^{1+2y} {}_{2}F_{1}(1+y,2+y,2,-e^{\tilde{\xi}})\\
\\
 &\xrightarrow{Y \,\gg\,1}& \sum^\infty_{n=0}\frac{(\Gamma(Y+n))^{2}}{(\Gamma(Y))^{2}}\frac{(-e^{\tilde{\xi}})^{n}}{n!(n+1)!}=e^{\tilde{\xi}}e^{-2Y(\ln Y-1)}\frac{1}{2Ye^{\frac{1}{2}\tilde{\xi}}}J_{1}(2Ye^{\frac{1}{2}\tilde{\xi}})
 \end{array}
\end{equation*}
and knowing the asymptotic representation of Bessel function we get
\begin{equation}
\label{I21LY}
\begin{array}{rcr}
I^{1}_{2}( Y, \tilde{\xi})&\xrightarrow{Y \,\gg\,1}& \frac{1}{2}e^{\tilde{\xi}}e^{-2Y(\ln Y-1)} \frac{1}{(Ye^{\frac{1}{2}\tilde{\xi}})^{\frac{3}{2}}}\left\lbrace \cos(\lbrace 8Ye^{\frac{\tilde{\xi}}{2}}-3\pi \rbrace/4)\left[ \frac{\Gamma(7/2)}{\Gamma(-1/2)}+O( (y^{2}e^{\tilde{\xi}})^{-1} )  \right]  \right.\\
\\
&&\left.-  \sin(\lbrace 8Ye^{\frac{\tilde{\xi}}{2}}-3\pi \rbrace/4)\left[ \frac{\Gamma(9/2)}{\Gamma(-3/2)}+O( (y^{2}e^{\tilde{\xi}})^{-1} )  \right]\right\rbrace
\end{array}
\end{equation}
Finally, we replace \eq{I22} by the integral to estimate the large $Y$ dependence of this term,   i.e.
    \beq \label{I22LY}
       I^2_2\Lb Y, \tilde{\xi}\Rb\,\,\to\,\,e^{\tilde{\xi}\,Y}\int ^Y_0d t\, \frac{\Gamma\Lb 1 + t - Y\Rb}{\Gamma\Lb 1 - t +Y\Rb\,\Gamma^2\Lb 1 + t\Rb}\,\,e^{-\tilde{\xi}\,t \,+ \,i\,\pi\,t}
    \eeq
   Taking the integral by the steepest decent method in the same way as in \eq{I21LY} we obtain the equation for the saddle point  $t_{SP}$:
 \beq \label{TSP1}
 - 2 \ln t_{SP} + 2 \ln\Lb Y - t_{SP} - 1\Rb\,-\,\tilde{\xi}\,=\,0\,; \,\,\,\,\,\,\,\,\,\,\,\,\,t_{SP} \,=\,\frac{Y -1}{1 + e^{\h\tilde{\xi}}}
\eeq
and
     \beq \label{I22SP}
      I^2_2\Lb Y, \tilde{\xi}\Rb\,\,\to\,\,\,\sqrt{\frac{\pi (Y - 1)e^{\h \tilde{\xi}} }{4 \Lb 1 - e^{\tilde{\xi}}\Rb}}\,e^{-2Y \ln Y + Y \tilde{\xi}\frac{e^{\h \tilde{\xi}}}{1 + e^{\h \tilde{\xi}}}}
      \eeq

\begin{figure}
 \begin{center}
  \leavevmode
      \includegraphics[width=8cm]{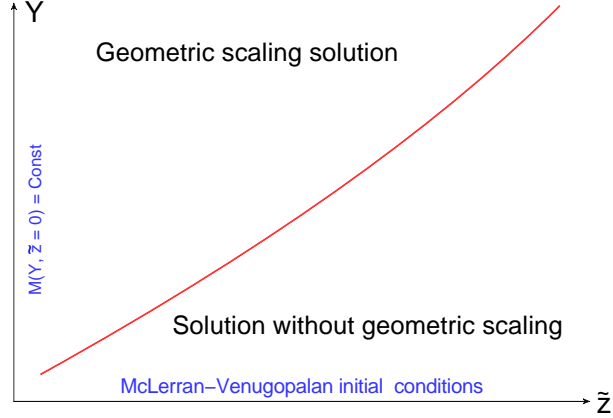}
      \end{center}
      \caption{Saturation region: two domains in which there is the geometric scaling behaviour of the solution and there is no such behaviour. The border red line is  \protect\eq{LINE}.}
      \label{line}
\end{figure}

    Comparing \eq{I11LY1}, \eq{I12SP}, \eq{I21LY} and  \eq{I22SP}  we see that at large values of $Y$  solutions of  \eq{BKMT7} and of \eq{YSOL} match each other on the line $\tilde{\xi}\,=\,- \,\ln Y$ which can be translate  as the line in $(Y, \tilde{z})$ plane (see \fig{line}):
    \beq \label{LINE}
    \tilde{z}\,\,=\,\,\bas \kappa Y\,\, + \,\,\ln Y
    \eeq

%%%%%%%%%%%%%%%%%%%%%%%%%%%%%%%%%%%%%%%%%%%%%%%%%%%%%%%%
 %%%%%%%%%%%%%%%%%%%%%%%%%%%%%%%%%%%%%%%%%%%%%%%%%%%%%%%%%
           %%%%%%%%%%%%%%%%%%%%%%%%%%%%%%%%%%%%%%
\begin{boldmath}
 \section{ Matching two solutions: at small  $\tilde{z}$ and at large $\tilde{z}$}
\end{boldmath}

 %%%%%%%%%%%%%%%%%%%%%%%%%%%%%%%%%%%%%%%%%%%%%%%%%%%%%%%%%% %%%%%%%%%%%%%%%%%
 %%%%%%%%%%%%%%%%%%%%%%%%%%%%%%%%%%%%%%
 \begin{boldmath}
 \subsection{Corrections at large $\tilde{z}$}
 \end{boldmath}

 %%%%%%%%%%%%%%%%%%%%%%%%%%%%%%%%%%%%%%%%%%%%%%%
 In this section we are going to find the first correction to the non-linear equation (see \eq{BKMR21})   deeply in the saturation region where we expect that solution has a geometric scaling behaviour, or in other words, it is a function of $\tilde{z}$.
One can see that the  equation for first correction takes the following form after substituting $M\,\,=\,\,M^{(0)}  + M^{(1)}$ into \eq{BKMR21}:
      \beq \label{MT4}
   \kappa \frac{d M^{(1)}\Lb \bar{z}\Rb}{d \tilde{z}} = \Big\{  \chi\Lb\gamma\Rb\,-\,\frac{1}{\gamma} \Big\}M^{(1)}\Lb \tilde{z}\Rb\,\,-\,\,\bar{z}  M^{(1)}\Lb \bar{z}\Rb \,\,+\,\, M^{(0)} \Lb \bar{z}\Rb \int^\infty_{\bar{z}}d z'  M^{(0)}\Lb z', Y\Rb
   \eeq
   where $\lambda = \int^\infty_0 M^{(0)}\Lb z'\Rb d z'\,\,=\,\,A \sqrt{\frac{\pi \kappa}{2}}$.

   In \eq{MT4} $M^{(0)}$ is the solutions to \eq{BKMR4} that takes the form
   \beq \label{MT5}
   M^{(0)}\Lb \bar{z}\Rb\,\,=\,\,A\,\exp\Lb - \bar{z}^2/\Lb 2 \kappa\Rb\Rb
   \eeq
   which we will use below. It  should be noted that  $A\,<\,0$ due to unitarity constraints; and \eq{BKMT7} for $A$ was derived from the matching of this solution at $\tilde{z}=0$.  Here, we wish to suggest a better procedure for matching.
   
   Using \eq{MT5} we can re-write the equation for the first correction as follows
     \beq \label{MT6}
   \kappa \frac{d M^{(1)}\Lb \bar{z}\Rb}{d \bar{z}} \,\,=\,\,\Big\{ \chi\Lb\gamma\Rb\,-\,\frac{1}{\gamma} \Big\}M^{(1)}\Lb \bar{z}\Rb\,\,-\,\, \bar{z}   M^{(1)}\Lb \bar{z}\Rb \,\,+\,\, M^{(0)} \Lb \bar{z}\Rb \int^\infty_{\bar{z}} d z' M^{(0)}\Lb z', Y\Rb
   \eeq
   After taking the integral the last term of this equation can be reduced to the form
   \beq \label{MT7}
   M^{(0)} \Lb \bar{z}\Rb \int^\infty_{\bar{z}}d z'  M^{(0)}\Lb z', Y\Rb  \,\,=\,\, M^{(0)} \Lb \bar{z}\Rb
   \,A\,\sqrt{\frac{\pi \kappa}{2}}\,\mbox{Erfc}\Lb \frac{\bar{z}}{\sqrt{2 \kappa}}\Rb  \,\,\xrightarrow{z \gg 1}\,\,A^2\,\frac{\kappa}{ \bar{z}} \,e^{ - \frac{\bar{z}^2}{ \kappa}}
   \eeq
   Using  the Mellin transform of \eq{MT}    we obtain  \eq{MT6} in the form:
   \beq \label{MT8}
  \Lb \kappa\,\gamma\, - \chi\Lb \gamma\Rb + \frac{1}{\gamma} \,+ \,\lambda \Rb m^{(1)}\Lb \gamma\Rb \,\,=\,\,\frac{ d m^{(1)}\Lb \gamma\Rb}{ d \gamma}\,\,+\,\,2\,A^2 \frac{\sqrt{\pi \kappa}}{\gamma} \,e^{\kappa \gamma^2/4}.
  \eeq
   The solution to \eq{MT8} takes the form (see Ref.\cite{MATH})
   \beq \label{MT9}
   m^{(1)}\Lb \gamma\Rb\,\,=\,\,m^{(0)}\Lb \gamma\Rb\,A^2\,\int^\infty_\gamma \frac{d \gamma'}{m^{(0)}\Lb \gamma'\Rb} \frac{\sqrt{\pi \kappa}}{\gamma'} \,e^{\kappa \gamma'^2/4}\,\,=\,\,m^{(0)}\Lb \gamma\Rb\,A^2\,\sqrt{ \pi \kappa}\,\int^\infty_\gamma \frac{d \gamma' }{\gamma'}\frac{\Gamma\Lb 1 - \gamma'\Rb}{\Gamma\Lb 1 + \gamma'\Rb} \,e^{ - \kappa \gamma'^2/4}
      \eeq
   where $m^{(0)}$ is given by \eq{BKMT3}.  As has been discussed in the integral of \eq{MT} we expect that large $\gamma$'s will be essential. In \eq{MT9} the typical $d \gamma' \sim 1/\sqrt{\kappa }\,\ll\,\gamma$ and we can replace this integral by
   \beq \label{MT10}
      m^{(1)}\Lb \gamma\Rb =  2 m^{(0)}\Lb \gamma\Rb\,A^2\,\sqrt{ \pi \kappa}\,\frac{\Gamma\Lb 1 - \gamma\Rb}{\Gamma\Lb 1 + \gamma\Rb}\int^\infty_\gamma \frac{d \gamma' }{\gamma'} \,e^{ - \kappa \gamma'^2/4} = - \,2\,m^{(0)}\Lb \gamma\Rb\,A^2\,\sqrt{ \pi \kappa}\,\frac{\Gamma\Lb 1 - \gamma\Rb}{\Gamma\Lb 1 + \gamma\Rb}\,
   Ei\Lb- \kappa \gamma^2/4\Rb
   \eeq
 See  Ref.\cite{RY}: formula {\bf 3.352(2)} for the last integration and formula {\bf 8.21} for the exponential integral $E\Lb x \Rb$.    For large $\gamma$ \eq{MT10} takes the form
 \beq \label{MT11}
       m^{(1)}\Lb \gamma\Rb\,\,=\,\,   \,m^{(0)}\Lb \gamma\Rb\,A^2\,\sqrt{ \pi \kappa}\,\,\frac{\Gamma\Lb 1 - \gamma\Rb}{\Gamma\Lb 1 + \gamma\Rb}\, \,  \frac{4}{\kappa \gamma^2}
   \exp\Big(- \kappa \gamma^2/4\Big)
  \eeq

    Plugging  \eq{MT11} in \eq{MT} we obtain $M^{(1)}\Lb \bar{z}\Rb$ in the form
    \beq \label{MT12}
    M^{(1)}\Lb \bar{z}\Rb\,\,\,=\,\,A^2\frac{4\,\sqrt{ \pi \kappa}}{\kappa}\,\,\int^{\epsilon + i \infty}_{ \epsilon - i \infty}\frac{d \gamma}{2 \pi i}\,\frac{1}{\gamma^2}\,e^{ \gamma \bar{z} \,\,+\,\,\frac{\kappa \gamma^2}{4}}\,\,=\,\,- A^2 e^{- \bar{z}^2/2 \kappa}\,D_{-2}\Lb \bar{z} \sqrt{\frac{2}{\kappa}}\Rb\,\,\xrightarrow{\bar{z} \gg 1} \,\,- A^2 \frac{\kappa}{\bar{z}^2} e^{ - \bar{z}^2/\kappa}
    \eeq
    where $D_p\Lb z\Rb$  is the parabolic cylinder function (see formulae {\bf 9.24 - 9.25 }  of Ref. [20]).
    
 %%%%%%%%%%%%%%%%%%%%%%%%%%%%%%%%%%%%%%%%%%%%%%%%%%%%%%%%%%%%%%%%%%%%%
 \begin{boldmath}
 \subsection{Corrections at small  $\tilde{z}$}
 \end{boldmath}

 %%%%%%%%%%%%%%%%%%%%%%%%%%%%%%%%%%%%%%%%%%%%%%%%%%%%%%%%%%%%%%%%%%%%%%%%
 In the region of small $\tilde{z}$ we can solve \eq{BKMR} for the amplitude $N\Lb \tilde{z}\Rb$  noting  that  the geometric scaling solution to the BFKL equation occurs at $\gamma = \gamma_{cr}$. In the vicinity of the saturation scale the linear BFKL equation can be simplified and replaced by
 \beq \label{MT130}
 \frac{d \widetilde{N}\Lb \tilde{z}\Rb}{ d \tilde{z}}\,\,=\,\,\Lb 1 - \gamma_{cr}\Rb \,\widetilde{N}\Lb \tilde{z}\Rb
  \eeq
  One can see that this equation leads to $ \widetilde{N}\Lb \tilde{z}\Rb\,\,\propto\,\,\exp\Lb ( 1 - \gamma_{cr})\tilde{z}\Rb$.  The  solution to the BFKL equation with full kernel in the vicinity of the saturation scale takes the form \cite{IIM,MUT,IIMU}
  \beq \label{MT131}
\widetilde{N}\Lb \tilde{z}; Q_s\Lb Y\Rb\Rb  \,\,=\,\,\phi_0\,e^{\Lb1 - \gamma_{cr}\Rb\tilde{z}}\exp\Bigg( - \frac{\tilde{z}^2}{2 \frac{\chi''_{\ga \ga}\Lb\gamma_{cr}\Rb}{\kappa}\,\ln \Lb Q^2_s\Lb Y\Rb/ Q^2_s\Lb 0\Rb\Rb}\Bigg)
\eeq 
  Therefore, we can trust \eq{MT130} only for 
  \beq \label{MT132}
  \tilde{z}\,~~\ll\, ~~ \sqrt{2 \frac{\chi''_{\ga \ga}\Lb\gamma_{cr}\Rb}{\kappa}\ln \Lb Q^2_s\Lb Y\Rb/ Q^2_s\Lb 0\Rb\Rb}\,\,\,\,=\,\,\,\,9.9\,\sqrt{ Y}
  \eeq
  
    For such $\tilde{z}$  the non-linear equation (see \eq{BKMR})  can be re-written in the form
 \beq \label{MT13}
   \kappa \, \frac{d \widetilde{N}\Lb \tilde{z}\Rb}{ d \tilde{z}}\,\,=\,\,\kappa \Lb 1 - \gamma_{cr}\Rb \,\widetilde{N}\Lb \tilde{z}\Rb\,\,-\,\, \widetilde{N}^2\Lb \tilde{z}\Rb
\eeq

The solution to this equation that satisfies the initial condition $\widetilde{N}\Lb \tilde{z} = 0\Rb = N_0$ takes the following form
\beq \label{MT14}
\widetilde{N}\Lb \tilde{z}  \Rb\,\,=\,\,\frac{ \kappa \Lb 1 - \gamma_{cr}\Rb\,N_0}{N_0 \,\,-\,\,\Lb N_0\,-\, \kappa \Lb 1 - \gamma_{cr}\Rb\Rb e^{ - \Lb 1 - \gamma_{cr}\Rb \,\tilde{z}}}
 \eeq
 
 As we will discuss in the next subsection the scattering amplitude in the coordinate space is equal to
 \beq \label{MT15} 
 N\Lb z \Rb \,\,=\,\,2 \frac{ d \widetilde{N}\Lb z  \Rb}{d z}\,\,=\,\,-2\, \frac{ \kappa \Lb 1 - \gamma_{cr}\Rb^2 \Lb N_0 - \kappa \Lb 1 - \gamma_{cr}\Rb \Rb \,N_0\, e^{ - \Lb 1 - \gamma_{cr}\Rb \,z}}{\Lb N_0 \,\,-\,\,\Lb N_0\,-\, \kappa \Lb 1 - \gamma_{cr}\Rb\Rb e^{ - \Lb 1 - \gamma_{cr}\Rb \,z}\Rb^2}
 \eeq
 We will use this equation in the matching procedure described below in the next section.

           %%%%%%%%%%%%%%%%%%%%%%%%%%%%%%%%%%%%%%

 \subsection{ Matching procedure}

 %%%%%%%%%%%%%%%%%%%%%%%%%%%%%%%%%%%%%%%%%%%%%%%%%%%%%%%%%%

In this subsection we would like to discuss matching of two  solutions at small and large $\tilde{z}$. First, we calculate the scattering amplitude of two dipoles in the coordinate representation where we know that at large $z$ this amplitude $ N\Lb z \Rb \to 1$.

FromÊ  \eq{MR} we see that

\beq \label{CR}
2 \pi \widetilde{N}\Lb  k_\bot, b; Y\Rb\,\,=\,\, 2\,\pi \int x d x  \frac{J_0\Lb k_\bot\, x\Rb}{x^2}N\Lb x, b;Y\Rb
\eeq
The main contribution to the integral over $x$ stems from $k_\bot\,x \leq 2.4$, where 2.4 is the zero of $J_0\Lb k_\bot x\Rb$ ($J_0\Lb k_\bot x = 2.4\Rb =0$), and therefore, we can replace the integral by
\beq \label{CR1}
 \widetilde{N}\Lb  k_\bot, b; Y\Rb\,\,=\,\h \,\int^{\frac{2.4^2}{k^2_\bot}}_0  \frac{d x^2}{x^2}N\Lb x, b;Y\Rb\,\,=\,\,\h \int^{\ln\Lb 2.4^2/k^2_\bot \Rb}_{- \infty} d \ln x^2 N\Lb x^2, b,Y\Rb
 \eeq
 or
 \beq \label{CR3}
 N\Lb x,b,Y\Rb \,\,=\,2\, \frac{d  \widetilde{N}\Lb  k_\bot, b; Y\Rb}{d \ln(2.4^2/k^2_\bot)}
 \eeq
 For the solution with the geometric scaling behaviour \eq{CR3} takes the form
 \beq \label{CR4}
 N\Lb z\Rb\,\,=\,\,2\, \frac{d \widetilde{N}\Lb \tilde{z}\Rb}{d \tilde{z}}
 \eeq
 
Bearing \eq{CR4} in mind we can formulate the matching procedure in the following way
   \bea \label{MATCHC}
   N_{z \,<\,1}\Lb z_m, N_0\Rb\,\,=\,\,   N^{(0)}_{z \,>\,1}\Lb z_m ,A\Rb\,\,=\,\,1 \,+\,2\,M^{(0)}\Lb z_m,A\Rb\nn\\
    \frac{d  N^{(0)}_{z \,<\,1}\Lb z_m, A\Rb}{d z}\,\,=\,\,+\,2\,\frac{d M^{(0)}\Lb z_m,A\Rb}{ d z}
   \eea
      %%%%%%%%%%%%%%%%%%%%%%%%%%%%%%%%%%%%%%%%%%%%%%%%%%%%%%%%%%
    \begin{figure}
 \begin{tabular}{ c c c}
  \leavevmode
      \includegraphics[width=5.5cm]{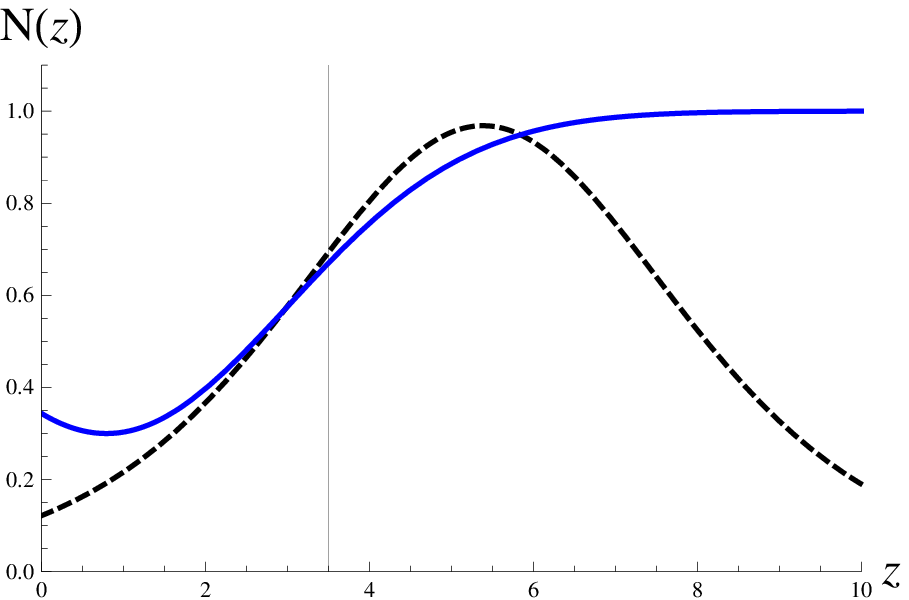}&      \includegraphics[width=5.5cm]{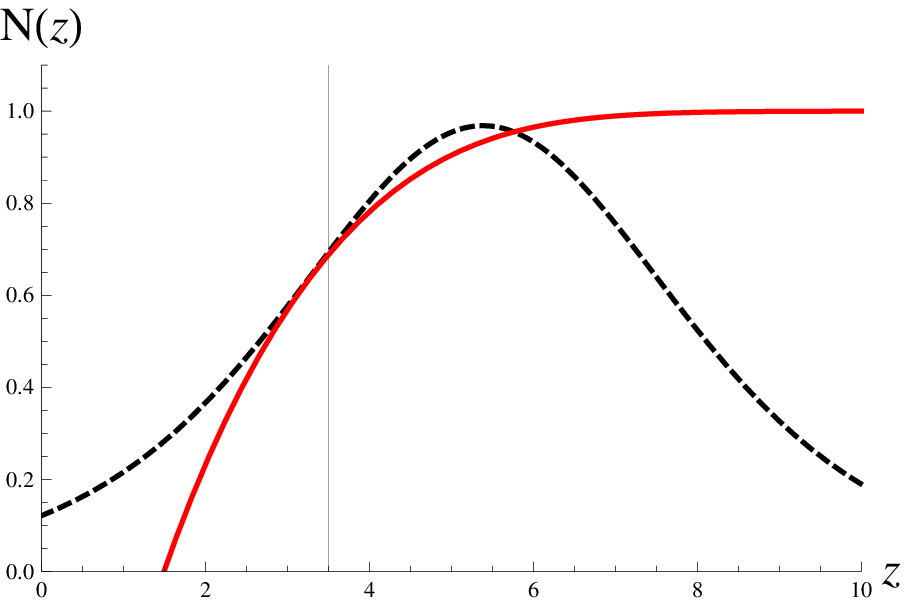}&    \includegraphics[width=5.5cm]{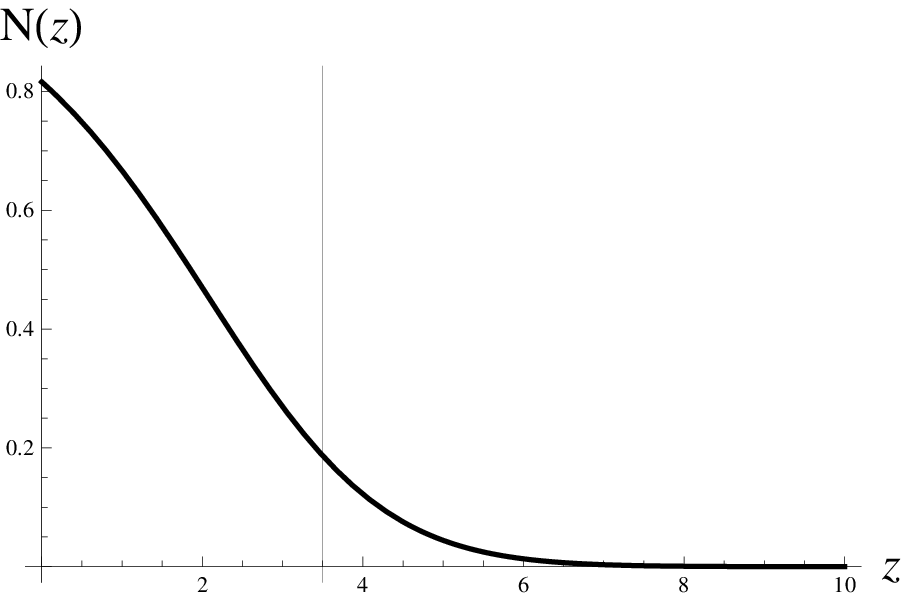}\\
      \fig{apr}-a &\fig{apr}-b & \fig{apr}-c\\
            \includegraphics[width=5.5cm]{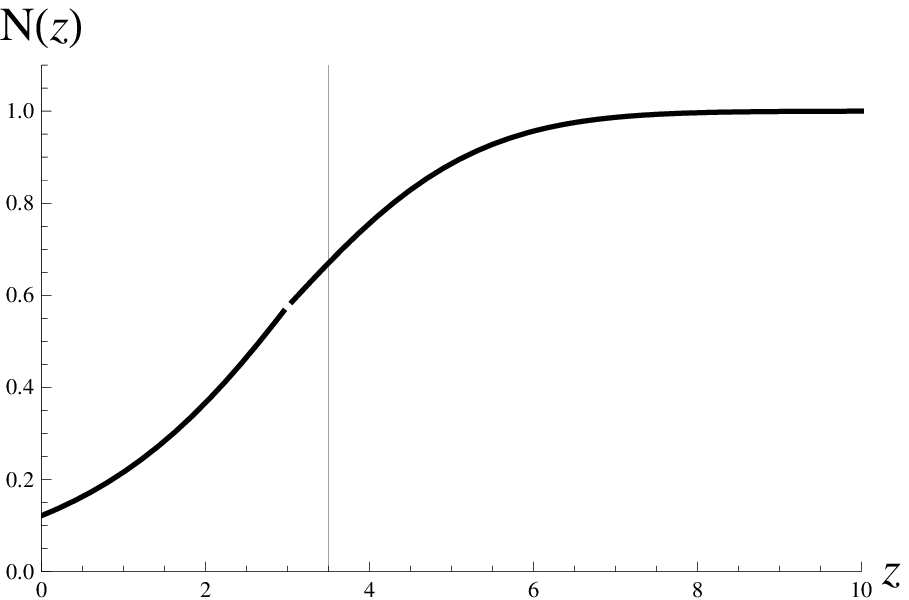}&      \includegraphics[width=5.5cm]{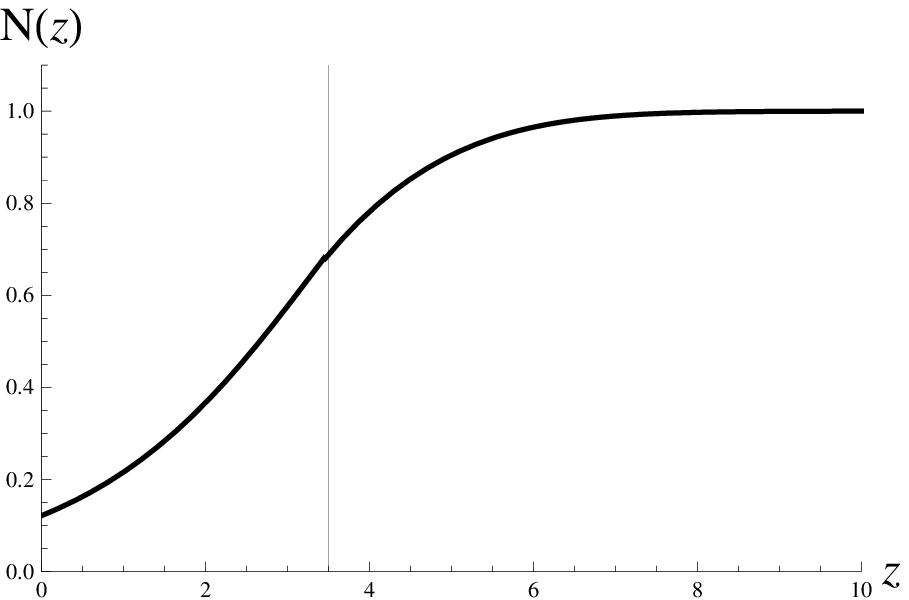}&    \includegraphics[width=5.5cm]{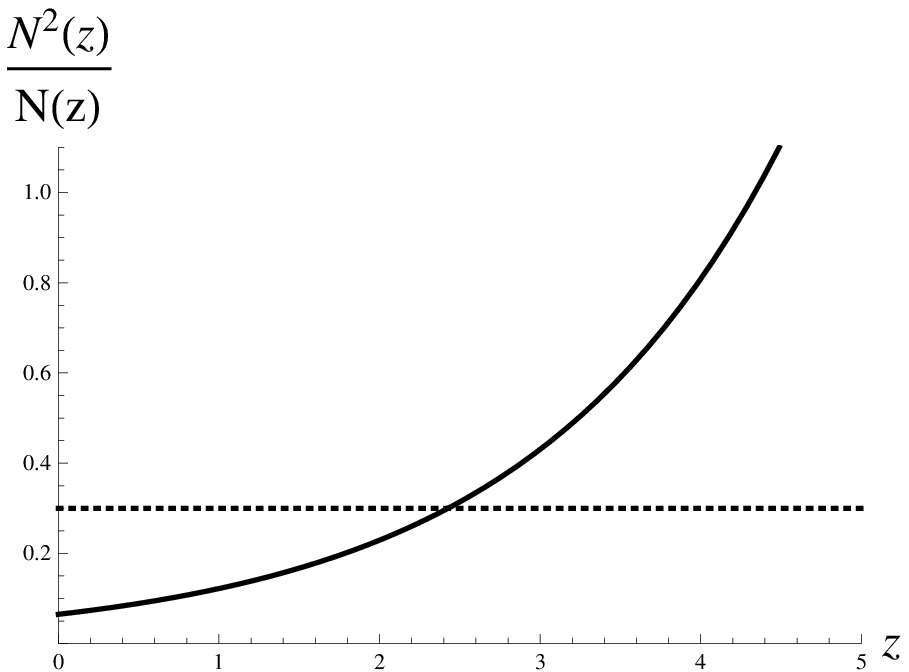}\\
      \fig{apr}-d  &\fig{apr}-e & \fig{apr}-f\\           \end{tabular}
      \caption{ The scattering amplitude $N\Lb z\Rb$  in the coordinate  representation: (\protect\fig{apr}-a)  matching of two solutions:  $N\Lb z\Rb$ at $z  < 1$ (  dashed  black b line) and  $N^{(0)}\Lb z \Rb z  \gg 1$( solid  blue  line) and (\protect\fig{apr}-b)  matching of two solutions:  $N\Lb z\Rb$ at $z  < 1$ (  dashed black line) and  $N^{(1)}\Lb z \Rb z  \gg 1$( solid red  line). The ratio $M^{(1)}\Lb z\Rb/M^{(0)}\Lb z\Rb$ is shown in \protect\fig{apr}-c. The value of $A$ is chosen  to be equal to 0.87 for \protect\fig{apr}-aÊ and $A$=0.65 for  \protect\fig{apr}-b.  The  vertical line marks $z_m = 3.5$ ,  for $z > z_m$ the corrections due to $M^{(1)}$ are small (less $30\protect\%$), while for $z < z_m$ the non-linear contributions to the amplitude  is less than $30\protect\%$ (see \protect\fig{apr}-f where the ratio of the contribution  proportional to $N^2\Lb z \Rb$ to the contribution proportional to $N\Lb z \Rb$ is shown). \protect\fig {apr}-d and \protect\fig{apr}-e show the resulting amplitude due to the matching procedure presented  in \protect\fig{apr}-a  and \protect\fig{apr}-b . The value of $N_0$ is chosen to be equal 0.1 in the picture.}
      \label{apr}
\end{figure}
 %%%%%%%%%%%%%%%%%%%%%%%%%%%%%%%%%%%%%%%%%%%%%%%%%%%%%%%%%
   One can see from \fig{apr}-a  that we cannot  find   the solution to both \eq{MATCHC}, but the second equation in \eq{MATCHC} is almost satisfied.  Actually, this matching supports the approach developed in Ref.\cite{IIMU}.
   From \fig{apr}-b one can see that the next to leading corrections at large $z$ drastically change the situation leading to the fact that both equations of \eq{MATCHC1}  are satisfied. 
   
      \bea \label{MATCHC1}
      N_{z\,<\,1}\Lb z_m, N_0\Rb\,\, &=&\,\,    N^{(1)}_{z\,\gg\,1}\Lb z_m, A\Rb\,\,=\,\,1\,+\, 2\,M^{(0)}\Lb z_m, A\Rb\,+\,2\,M^{(1)}\Lb z_m, A\Rb  ; \nn\\
      \frac{d  N_{z \,<\,1}\Lb z_m, N_0\Rb}{d z}\,\,&=&\,\frac{d  N^{(0)}_{z \,\gg\,1}\Lb z_m, A\Rb}{d z}\,\,=\,\,\,\,2\,\frac{d \Lb M^{(0)}\Lb z_m, A\Rb\,+\,M^{(1)}\Lb z_m, A\Rb\Rb}{ d z}
   \eea
   It is interesting to note that  the solution to \eq{MATCHC1}: $z_m = 3.5$ and $ A=0.65$, gives such  values  of these parameters that matching occurs in the region where the next-to-leading corrections to  asymptotical contribution ($M^{(1)}$) is less or about of 30\% ($M^{(1)}\Lb z_m\Rb/M^{(0)}\Lb z_m\Rb \leq 0.3$). On the other hand the non-linear corrections are not large ($< $30\%) and we can trust the simplified \eq{MT13}, since $z_m$ satisfies \eq{MT131} even at $Y=1$.
   
    In general we consider this matching as a strong argument for the procedure suggested in   Rev.\cite{IIMU} for finding the approximate solution of the non-linear equation which is suited for phenomenology.
   
   \section{Conclusions}
   In this paper we show that at large $z$ the solution to Balitsky-Kovchegov equation  takes the following form
   \beq
   N\Lb z \Rb\,\,=\,\,1 - A \exp\Bigg( - \frac{\Lb z - A \sqrt{\kappa \pi/2} + 2\psi(1)\Rb^2}{	 2 \kappa}\Bigg)
   \eeq
   which is the same as solution given in Ref.\cite{LT} at $z\, \gg \,1$. However,   the asymptotic behaviour of the solution depends on different variable $\bar{z} \,= \, z - A \sqrt{\kappa \pi/2} + 2\psi(1 )$ while  the solution at small $z$ in the vicinity of the saturation scale is determined by $z$. This observation, we believe, is essential for understanding the matching of the solutions at small and large $z$ and for searching the solution for intermediate $z$.
   
   We found the solution in the entire kinematic region at large $z$ which satisfies the  McLerran-Venugopalan initial condition. This problem has been discussed in Refs.\cite{LTHI,KLT}  in the case of simplified kernels but here we give the solution for the full BFKL kernel.
   
   The next-to-leading in the region of large $z$  has been calculated and it is demonstrated that this correction change crucially the matching with the solution in the vicinity of the saturation scale.
   
   We hope that this paper will be useful  for finding general features of the behaviour of the dipole scattering amplitude in the saturation region.
              %%%%%%%%%%%%%%%%%%%%%%%%%%%%%%%%%%%%%%%%%%%%%%%%%%%%
      \section{Acknowledgements}
  
    %%%%%%%%%%%%%%%%%%%%%%%%%%%%%%%%%%%%%%%%%%%%%%%%%%%%%%%%%
 We thank our    colleagues at UTFSM and Tel Aviv university for encouraging discussions.   This research was supported by the BSF grant 2012124  and by the  Fondecyt (Chile) grants  1140842  and 1120920  and    DGIP USM  grant  11.13.12  .

     \end{document}